\documentclass{jps-cp}
\usepackage{txfonts} 
\title{Nucleon Structure from Lattice QCD Calculations}

\author{Jian-Wei \textsc{Qiu}$^{1}$}

\inst{$^{1}$Theory Center, Jefferson Lab, 12000 Jefferson Avenue, Newport News, VA 23606, U.S.A.
}

\email{jqiu@jlab.org}

\recdate{February 15, 2019}

\abst{Parton distribution and correlation functions describe the relation between a hadron and the quarks and gluons (or collectively, the partons) within it, and carry rich information on hadron's partonic structure that cannot be calculated by QCD perturbation theory.  In this talk, I will review what lattice QCD can and cannot do for calculating the parton distribution and correlation functions, and the new ideas and efforts around the world to explore nucleon structure from lattice QCD calculations by combining the strength of both lattice QCD and perturbative QCD in such a way that is complementary to our on-going effort to extract these fundamental functions from experimental data.
}

\kword{Nucleon Structure, Lattice QCD, Perturbative QCD, Factorization}

\begin{document}
\maketitle

\section{Introduction}
\label{sec:intro}

The proton and neutron (or collectively, known as the nucleon) are the fundamental building blocks of all atomic nuclei and make up essentially all the visible matter in the universe, including the stars, the planets, and us. The nucleon emerges as a strongly interacting, relativistic bound state of quarks and gluons in Quantum Chromodynamics (QCD), and has a complex internal structure only beginning to be revealed in modern experiments and lattice QCD calculations. Both theory and technology have now reached a point where human is capable of exploring the inner dynamics and structure of nucleons and nuclei at the sub-femtometer distance, which provides new opportunities for us to understand ourselves and the visible world around us in a deeper and more fundamental level and is expected to lead to a new emerging science of nuclear femtography.  

QCD, a dynamical theory of quarks and gluons, is believed to be the theory of the strong interaction physics and is responsible for the known macroscopic properties of nucleons, such as mass and spin, as well as their rich, mysterious and largely unknown microscopic internal structure in terms of the properties and interactions of quarks and gluons in QCD.  With its {\it color confinement} - the defining property of QCD, no modern detector has ever seen quarks and gluons in isolation, and it has been an unprecedented intellectual challenge to explore and to quantify the nucleon's internal structure without being able to see quarks and gluons directly.  

Unlike the well-known atomic, molecular or crystal structure, which is often presented in terms of a ``still picture'' showing the relative location of various nuclei, the internal structure of nucleon cannot be described by any ``still picture''.  This is because nuclei are so small comparing to the size of atom, and so heavy that their motion is so much slower than the speed of light, while the motion of quarks and gluons is fully relativistic, their numbers are constantly changing due to quantum fluctuations inside the hadron, and their color is quantum mechanically entangled.  Consequently, we quantify the nucleon's partonic structure in QCD in terms of ``quantum probabilities'' to find quarks, gluons and their correlations, which are defined in terms of hadron matrix elements of quark ($\overline{\psi}$, $\psi$) and gluon ($A^{\mu}$) operators: $\langle P,S|{\cal O}(\overline{\psi},\psi,A)|P,S\rangle$ with hadron momentum $P$ and spin $S$.  Although these ``quantum probabilities'' or hadron matrix elements are well defined in QCD, none of them is a direct physical observable, exactly due to the {\it color confinement} and the fact that no quarks and gluons could be observed directly in a physical detector. That is, we have to identify ``probe(s)'' that can ``see'' the quarks and gluons inside the nucleon in order to measure these quantum probabilities to quantify the nucleon's partonic structure or internal landscape.

Fortunately, QCD has another equally important and fundamental defining property, the {\it asymptotic freedom} - the strong force effectively is weak if it is probed at a sufficiently short-distance.  It is the asymptotic freedom that makes it possible for us to develop the powerful theoretical formalism, known as QCD factorization~\cite{Collins:1989gx} that factorizes a physical cross section with measured hadron(s) in high energy collision into a product of a calculable partonic scattering of quark(s) and/or gluon(s) at a sub-femtometer scale and a set of fundamental and universal matching functions to link the scattering quark(s) and/or gluon(s) to the identified hadron(s) within well-defined approximations.  For example, inclusive cross section of electron ($e$) - proton ($p$) deeply inelastic scattering (DIS) has one identified hadron - the proton of momentum $p$ and could be factorized as,
\begin{equation}
\sigma^{\rm DIS}_{ep\to e'X}(x_B,Q^2) = \sum_{i=q,\bar{q},g} \int_{x_B}^{1}\frac{dx}{x}\, 
\hat{\sigma}_{ei\to e'X'}\left(\frac{x_B}{x},\frac{Q^2}{\mu^2};\alpha_s(\mu^2)\right) f_{i/p}(x,\mu^2)
+{\cal O}\left(\frac{1}{(QR)^2}\right)\, ,
\label{eq:dis}
\end{equation}
where $x_B=Q^2/2p\cdot q$ and $Q^2=-q^2$ with the $q$ as the momentum transfer between the scattering electron and proton, and $R \sim 1/\Lambda_{\rm QCD} \sim 1~\text{fm}$ is the hadron radius representing the order of nonperturbative hadronic scale.   The factorization formalism in Eq.~(\ref{eq:dis}) presents an indirect and approximate way to probe quarks and gluons in electron-proton scattering when the momentum transfer $Q^2\gg1/R^2$, in terms of the ``probe'': $\hat{\sigma}_{ei\to e'X'}$ - a perturbatively calculable electron scattering off a {\it parton} of flavor $i=q,\bar{q},g$ (quark, antiquark, gluon, respectively), the universal matching functions: $f_{i/p}(x,\mu^2)$ - the probability distributions to find the {\it parton} of flavor $i$ in the proton $p$ carrying its momentum fraction between $x$ and $x+dx$, probed at the factorization scale $\mu$, and a controlled ``correction'' suppressed by the power of the large momentum transfer $Q$.  It is the precision, for which we can calculate the $\hat{\sigma}_{ei\to e'X'}$ in QCD perturbation theory at the scale $Q$ - a consequence of the {\it asymptotic freedom}, that provides us the well-controlled short-distance ``probe(s)'' to effectively ``see'' the quark(s) and/or gluon(s) at a sub-femtometer distance ($\sim 1/Q$); it is the matching functions that link the quark and/or gluon at such a short distance inside the observed hadron to provide the rich information on the hadron's partonic structure; and it is the universality of these matching functions, extracted from one or few experimental measurements and then used to predict and to be tested in many other experiments, that ensures the predictive power of QCD.   With the factorization formalisms available for various observables, QCD has been extremely successful in interpreting all available data from high energy scatterings with probing distance less than 0.1~fm (or equivalently, with the momentum transfer in the scattering larger than 2~GeV), which has provided us the confidence and the tools to discover the Higgs particle and to explore new physics beyond the Standard Model of particle physics in high energy hadronic collisions~\cite{Brambilla:2014jmp}.  

The predictive power of QCD factorization formalisms, like the one in Eq.~(\ref{eq:dis}), relies on the precision of these universal matching functions, which are defined in terms of hadronic matrix elements of quark-gluon operators and encoded with rich information on hadron's partonic structure.  The $f_{i/p}(x,\mu^2)$ in Eq.~(\ref{eq:dis}), known as the parton distribution functions (PDFs), are defined as
\begin{equation}
f_{q/p}(x,\mu^2) = \int \frac{dp^+\xi^-}{2\pi} e^{-ixp^+\xi^-} 
\langle p| \overline{\psi}_q(\xi^-)\frac{\gamma^+}{2p^+}\exp\left\{-ig\int_0^{\xi^-} d\eta^- A^+(\eta^-)\right\} \psi_q(0)|p\rangle
\label{eq:pdf}
\end{equation}
where the light-cone components: $v^{\pm}\equiv (v^0\pm v^3)/\sqrt{2}$ are defined for any four-vector $v^\mu=(v^0,\mathbf{v}_\perp, v^3)$, and might be the most fundamental and the best studied matching functions. In the framework of QCD factorization, enormous efforts have been devoted to determine PDFs and their uncertainties in terms QCD global analysis of all existing high energy scattering data by many collaborations around the world, including MMHT~\cite{Harland-Lang:2014zoa}, CT~\cite{Dulat:2015mca}, NNPDF~\cite{Ball:2017nwa},  HERAPDF~\cite{Alekhin:2017kpj}, and JAM~\cite{Ethier:2017zbq}.  Although we have improved our knowledge of PDFs tremendously, they still have large uncertainties in both large and small $x$ region, and sometimes, the uncertainties of PDFs become the main sources of systematic errors for comparison between the best theoretical calculations available and the better and more precise experimental data from the LHC.  With the hadronic matrix elements well-defined in QCD, it is not only very natural, but also critically important to ask if we can calculate these PDFs and other partonic correlation functions directly in QCD, or at least in Lattice QCD (LQCD), which is by far the most reliable and theoretically justified approach to calculate the nonperturbative quantities in QCD.  Unfortunately, it is, if not impossible, very difficult to calculate PDFs and other partonic correlation functions directly in LQCD because the partonic operators of these hadron matrix elements live on the light-cone, as shown in Eq.~(\ref{eq:pdf}), and are time-dependent with the time defined in Minkowski space, while LQCD calculations are performed in Euclidean space with the Minkowski time $t$ analytically continued to a Euclidean time, $\tau = i\, t$.

LQCD calculation of hadronic matrix elements is based on the QCD path integral formalism evaluated in a discretized finite volume of Euclidean space-time with the integration of all field configurations done by the Monte Carlo method~\cite{Lin:2017snn}.  It requires a minimum input to define the QCD action, such as the overall hadronic scale and quark masses, which could be fixed by a few known baryon and meson masses.  Due to the computation resources and technical challenges, early LQCD efforts for determining hadron's partonic structure were limited to calculations of {\it local} matrix elements.  With the improvement of computation power and remarkable new ideas in recent years, there has been a tremendous growth of LQCD efforts for calculating {\it nonlocal} matrix elements, from which parton distribution and correlation functions could be extracted~\cite{Lin:2017snn,Cichy:2018mum}.  In the rest of this writeup for my talk, I will review briefly what lattice QCD can and cannot do for calculating the parton distribution and correlation functions that provide rich information on hadron's partonic structure.  I will first briefly summarize the early LQCD effort to study hadron structure in terms of the calculation of local matrix elements in the next section.  In the Sec.~\ref{sec:nonlocal}, I will focus on the fast developing world effort for extracting the parton distribution and correlation functions by calculating non-local matrix elements in LQCD, and finally, I will provide the summary and outlook in the Sec.~\ref{sec:summary}.

\section{Hadron structure in terms of local matrix elements}
\label{sec:local}

Hadron structure with relativistic quarks and gluons is dynamical and nonlocal.  Parton distribution and correlation functions, which carry the structure information, cannot be directly calculated in LQCD defined with a Euclidean space-time.  Historically, LQCD has been used to calculate hadron form factors of local currents~\cite{Alexandrou:2018sjm} and Mellin moments of PDFs, which are given in terms of hadronic matrix elements of local twist-2 operators of quark and gluon fields~\cite{Lin:2017snn}.  

\begin{figure}[tbh]
\centering
\begin{minipage}[c]{0.4\textwidth}
\centering
\includegraphics[width=0.9\textwidth,angle=270]{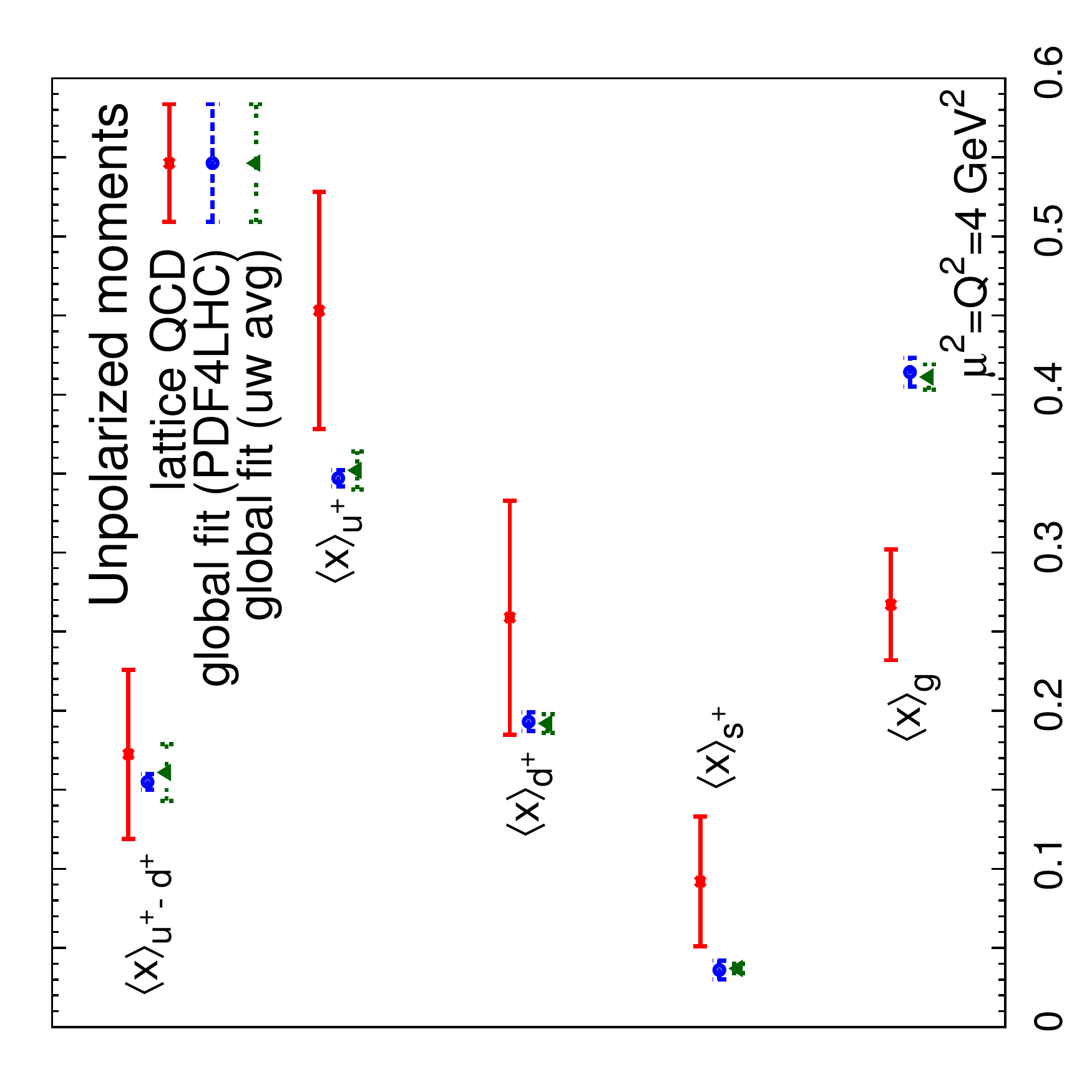}
\end{minipage}
\hskip 0.2in
\begin{minipage}[c]{0.4\textwidth}
\centering
\includegraphics[width=0.9\textwidth,angle=270]{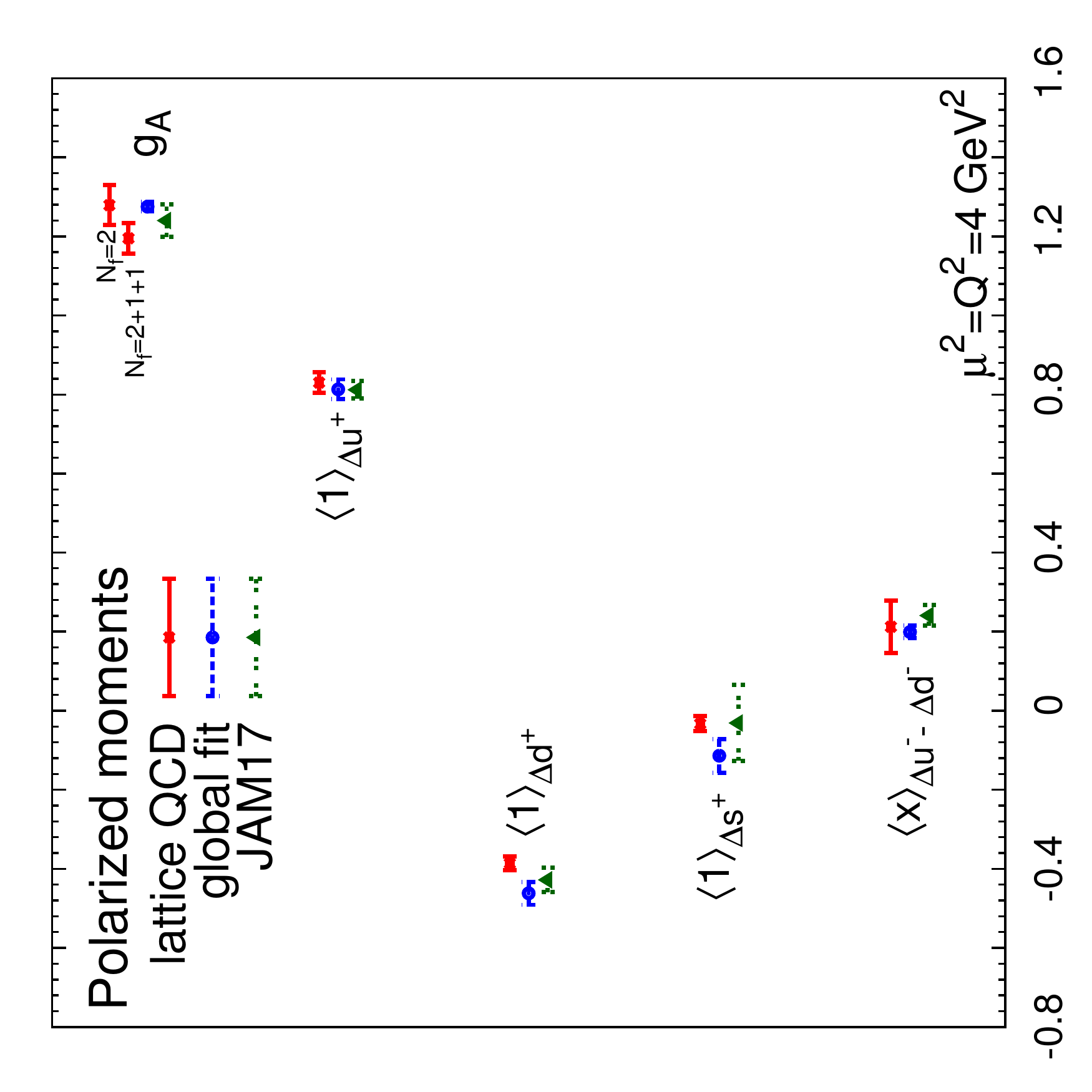}
\end{minipage}
\caption{A comparison of the Mellin moments of PDFs determined by the lattice QCD computations 
and QCD global fits:  unpolarized (Left) and polarized (Right)~\cite{Lin:2017snn}.
}
\vspace{-0.2in}
\label{fig:moments}
\end{figure} 
In principle, we could determine the nonlocal correlation of quark and gluon fields and reconstruct PDFs if we are able to calculate a sufficient number of the moments in LQCD.  The Mellin moments of PDFs are defined as
\begin{equation}
\langle x^n \rangle_f(\mu^2) \equiv \int_0^1 dx\, x^n f(x,\mu^2)\, ,
\quad \quad
f=\{u,\bar{u},d,\bar{d},s,\bar{s}, ...; g\}\, ,
\label{eq:moments}
\end{equation}
with $q^{\pm}\equiv q \pm \bar{q}$ for quark flavors and $\Delta f$ for helicity distribution functions.  LQCD calculation has done a good job in determining the lower moments of PDFs.  In Fig.~\ref{fig:moments}, we compare the lower moments of PDFs determined by QCD global fits and LQCD calculations.  In the left plot, we show the first moment of unpolarized PDFs, which is equal to the momentum fraction carried by the quark and gluon; and in the right plot, we present the zeroth moment of quark helicity distributions, which is equal to the helicity fraction of a fast moving proton carried by the quarks.   With the precise date from experiments and a wide range of collision energies, QCD global fits have provided much accurate determination of the momentum fractions carried by quarks and gluons, which effectively set the benchmarks for LQCD calculations.  On the other hand, LQCD determination of quark helicity contribution to the proton spin is comparable to what have been determined by QCD global fits.  With less data available for providing information on quark transversity distribution, precise determination of quark tensor charge from LQCD calculations has helped to better determine the $x$-dependence of quark transversity distribution~\cite{Lin:2017stx}.

However, due to the power divergent mixing between twist-2 operators on the finite lattice, it is extremely difficult and effectively unpractical to determine the higher moments of PDFs. In practice, LQCD calculations of the Mellin moments of unpolarized PDFs are limited to the lowest three moments with the first moment results in Fig.~\ref{fig:moments}, and the other two moments in Refs.~\cite{Dolgov:2002zm,Gockeler:2004wp}, respectively.  With the limited information on the lowest three moments, it is not realistic to determine the $x$-dependent PDFs to be comparable with those from QCD global fits~\cite{Detmold:2001dv}.

\section{Hadron structure from non-local matrix elements}
\label{sec:nonlocal}

Over the years, various ideas have been proposed to extract $x$-dependent PDFs from LQCD calculations, such as the path-integral formulation of the deep-inelastic scattering hadronic tensor~\cite{Liu:1993cv,Liu:1999ak} and the inversion method~\cite{Horsley:2012pz}.   Recently, Ji~\cite{Ji:2013dva} introduced a set of LQCD-calculable quasi-PDFs,
\begin{equation}
\tilde{q}(\tilde{x},\mu^2,P_z) = \int \frac{dP_z\xi_z}{2\pi} e^{-i\tilde{x}P_z\xi_z} 
\langle P| \overline{\psi}_q(\frac{\xi_z}{2})\frac{\gamma_z}{2P_z}
\exp\left\{-ig\int_0^{\xi_z} d\eta_z A_z(\eta_z)\right\} \psi_q(\frac{-\xi_z}{2})|P\rangle
\label{eq:qpdf}
\end{equation}
for quasi-quark distribution, and argued that the quasi-PDFs of hadron momentum $P_z$ become corresponding PDFs when $P_z$ is boosted to infinity. Recognizing the fact that a separation in $\xi_z$ at an equal time becomes a separation in $\xi^-$ while $\xi^+=0$ when $P_z\to \infty$, and the idea to calculate hadron matrix elements of parton correlation along the $z$-direction in LQCD is novel.  Taking the limit, $P_z\to \infty$ is difficult to achieve in LQCD calculation due to the limited size of lattice spacing.  A perturbative matching of quasi-PDFs at a finite $P_z$ to the standard light-cone PDFs was proposed~\cite{Ji:2013dva}, 
\begin{equation}
\tilde{q}(\tilde{x},\mu^2,P_z) = \int_{\tilde{x}}^1 \frac{dy}{y} Z\left(\frac{\tilde{x}}{y},\frac{\mu}{P_z}\right)q(y,\mu^2) 
+{\cal O}\left(\frac{\Lambda^2}{P_z^2},\frac{M^2}{P_z^2}\right) 
\end{equation}
with $q(y,\mu^2)$ being the standard light-cone quark distribution and perturbatively calculable matching coefficient, $Z$ and power corrections in $1/P_z$.  It was noted recently that instead of $1/P_z^2$, the power corrections could be proportional to $1/[x^2(1-x)P_z^2]$, which could be significantly enhanced at a large $x$ or small $x$~\cite{Braun:2018brg}.  Comparing Eqs.~(\ref{eq:pdf}) and (\ref{eq:qpdf}), one recognizes that the operator defining the quasi-quark distribution is not a twist-2 operator, and consequently, unlike the quark distribution, quasi-quark distribution is not boost invariant and its dimension-three operator leads to a ultraviolet (UV) linear power divergence, which makes the renormalization of the LQCD calculation much more challenging~\cite{Alexandrou:2017huk,Stewart:2017tvs}. 

Regardless the difficulties, tremendous efforts have been investigated by both LQCD and perturbative QCD (PQCD) communities to study the quasi-PDFs in recent years, and important progresses have been made in our understanding of the relationship between the PDFs and quasi-PDFs, and their determination from LQCD~\cite{Lin:2017snn,Cichy:2018mum,Alexandrou:2019lfo}.  The power UV divergence of quasi-quark distributions were proved to be multiplicative renormalizable by three groups~\cite{Ji:2017oey,Ishikawa:2017faj,Green:2017xeu}.  While the operator defining the quasi-gluon distribution has a dimension of four, QCD color gauge invariance ensures that its maximum power of UV divergence is linear, which were proved to be multiplicative renormalizable by two groups with completely different methods~\cite{Zhang:2018diq,Li:2018tpe}. With the proof of QCD collinear factorization of hadronic matrix elements of operators defining quasi-PDFs in terms of PDFs~\cite{Ma:2014jla}, extracting PDFs from LQCD calculated hadronic matrix elements of quasi-parton operators has a solid theoretical foundation, while technical challenges, such as nonperturbative renormalization of power divergence and large hadron momentum in LQCD calculations, still exist.

Meanwhile, a number of other approaches to extract PDFs from LQCD calculations were also proposed, such as the good ``lattice cross section'' - a QCD factorization based general approach to calculate PDFs in LQCD {\it indirectly}~\cite{Ma:2014jla,Ma:2017pxb}, the pseudo-PDFs~\cite{Radyushkin:2017cyf,Orginos:2017kos}, and the ``OPE without OPE'' approach~\cite{Chambers:2017dov}.
The good ``lattice cross section'' (LCS) is defined as single hadron matrix elements that satisfy (1) calculable in Euclidean-space LQCD, (2) renormalizable for UV divergences to ensure a reliable continue limit, and (3) factorizable to PDFs with infrared-safe matching coefficients. It is the (3)-factorization that relates the desired PDFs to the LQCD calculable LCSs, just like how PDFs are related to the factorizable hadronic cross sections.  To overcome the power UV divergence of the operator defining the quasi-PDFs, a set of current-current correlators were introduced for good LCSs~\cite{Ma:2017pxb}. With renormalized or conserved currents, the UV renormalization of the good LCSs is trivial; and with many choices and combinations of currents, this approach could help construct many more LQCD calculable quantities to provide more information and constraints to determine the hadron's partonic structure.  Similar to the extraction of PDFs from experimental data of factorizable and measurable hadronic cross sections, PDFs could be extracted from QCD global analysis of data generated by LQCD calculations of good LCSs.   

\vspace{-0.15in}
\begin{figure}[!h]
\centering
\begin{minipage}[c]{0.46\textwidth}
\includegraphics[width=0.96\textwidth,height=1.6in]{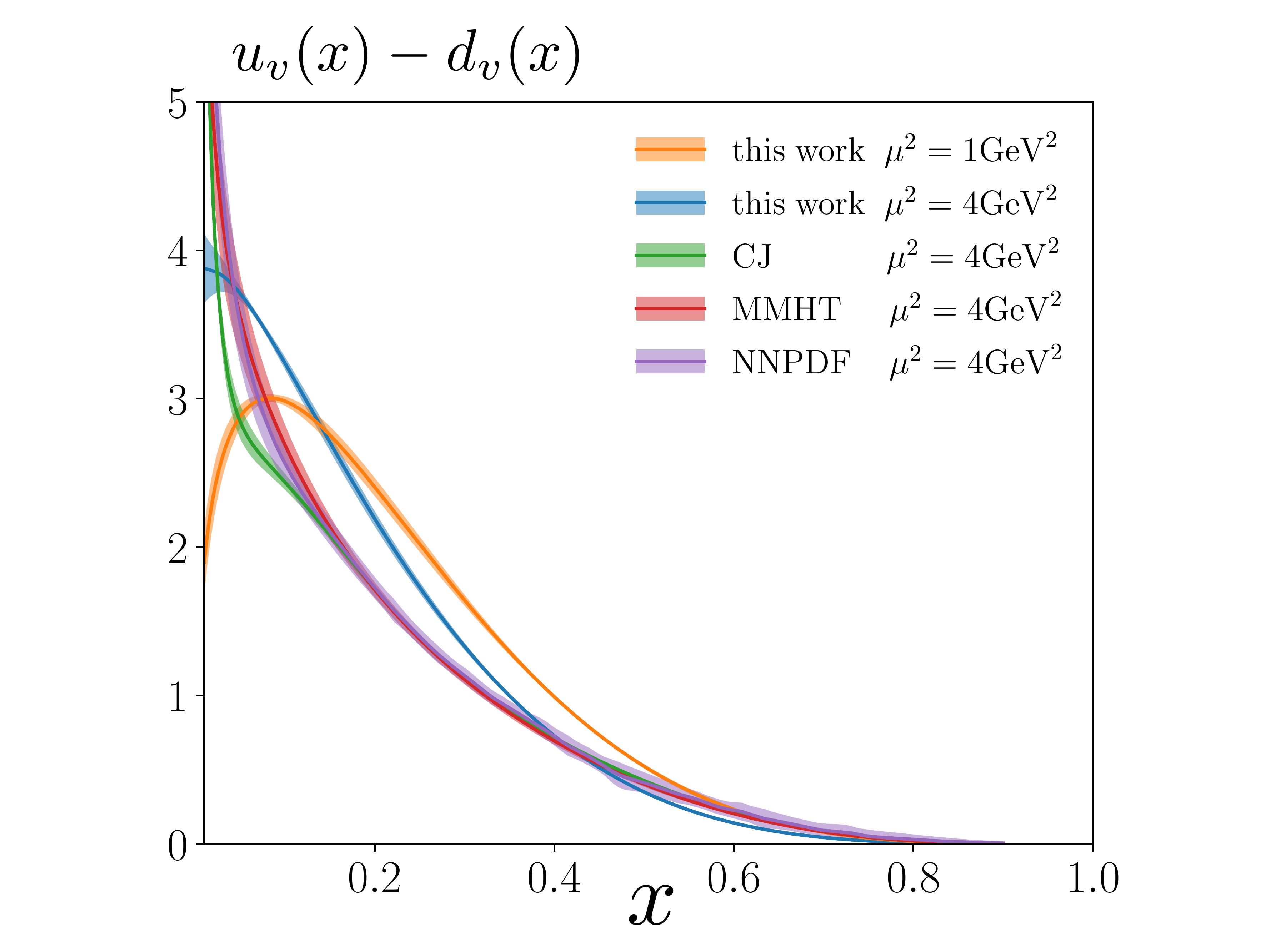}
\end{minipage}
\hskip 0.10in
\begin{minipage}[c]{0.4\textwidth}
\centering
\includegraphics[width=0.9\textwidth,height=1.6in]{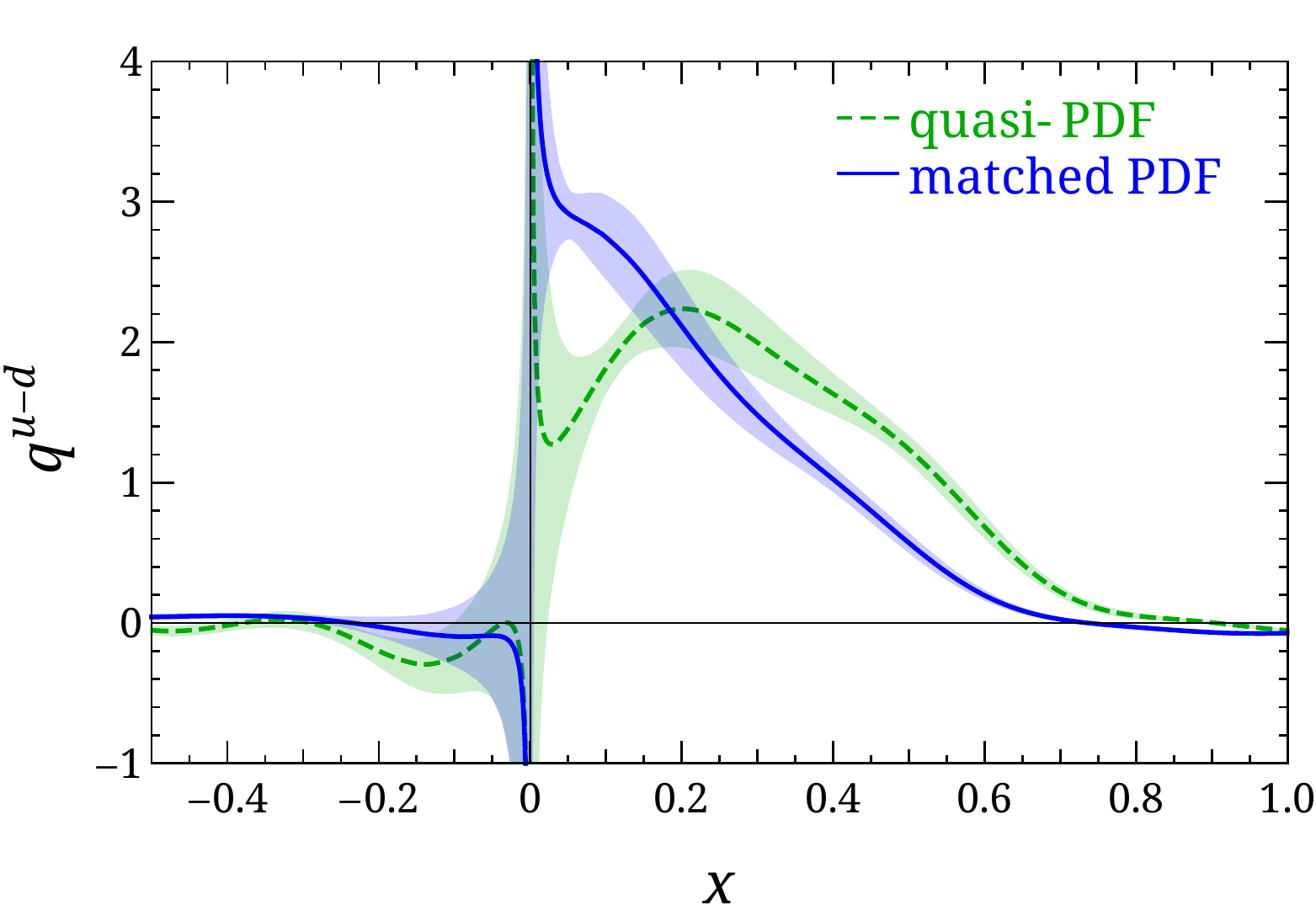}
\end{minipage}
\caption{Left:\ Scale dependence of Pseudo-quark distribution function~\cite{Orginos:2017kos}. Right:\ Effect of one-loop matching between quasi-quark and quark distribution function~\cite{Chen:2018xof}.  
}
\label{fig:pseudo}
\end{figure} 
\vspace{-0.1in}

Pseudo-PDFs approach was introduced to better control the power UV divergence of quasi-PDFs in LQCD calculation, and to avoid the operator mixing of the quasi-quark distribution under the renormalization~\cite{Orginos:2017kos}. A pseudo-quark distribution is defined as,
\begin{equation}
{\cal P}(x,\xi^2)\equiv \int \frac{d\omega}{2\pi} e^{-ix\omega} {\cal M}_{p=p^0}(\omega,\xi^2)/{\cal M}_{p=p^0}(0,\xi^2)
\label{eq:pseudo}
\end{equation}
where the hadronic matrix element ${\cal M}_{p}(\omega,\xi^2)$ with the Ioffe time $\omega=p\cdot \xi$ is defined as 
\begin{equation}
{\cal M}^\alpha(\omega,\xi^2)\equiv
\langle p|\overline{\psi}(\xi)\gamma^\alpha \exp\left\{-ig\int_0^{\xi} d\lambda v\cdot A(\lambda v)\right\} \psi(0)|p\rangle
\approx 2p^\alpha {\cal M}_p(\omega,\xi^2)\, .
\label{eq:matrix}
\end{equation}
With the ratio in Eq.~(\ref{eq:pseudo}), the power UV divergence of the matrix elements cancels since the UV divergence is multiplicatively renormalizable and is insensitive to the details of the long-distance hadron states.  The $\xi^2$ is the hard scale at which the distribution is probed and its dependence should obey the DGLAP evolution, which is clearly demonstrated in the left plot in Fig.~\ref{fig:pseudo}. 

\vspace{-0.15in}
\begin{figure}[!h]
\centering
\begin{minipage}[c]{0.4\textwidth}
\centering
\includegraphics[width=0.92\textwidth]{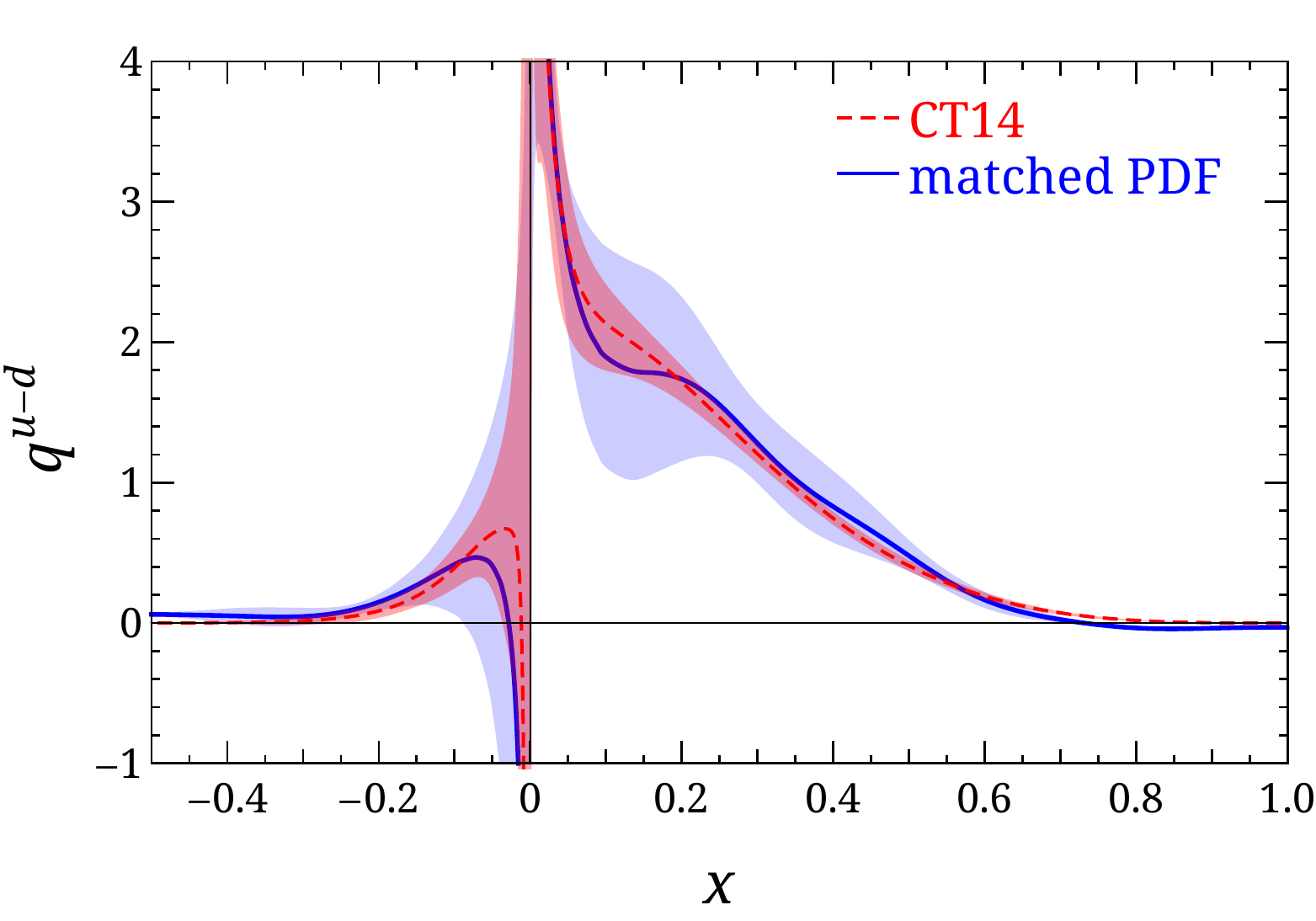}
\end{minipage}
\hskip 0.2in
\begin{minipage}[c]{0.4\textwidth}
\centering
\includegraphics[width=0.88\textwidth]{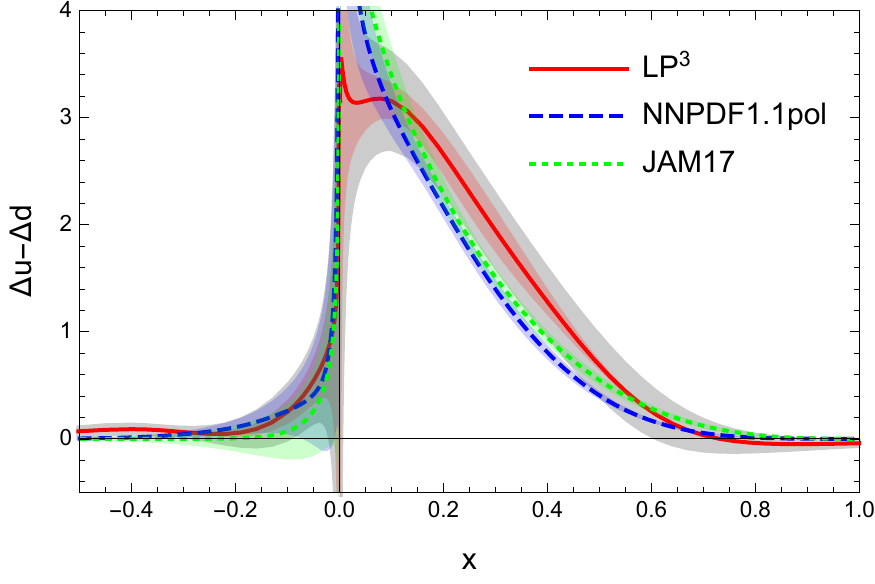}
\end{minipage}
\caption{LQCD determination of $u(x)-d(x)$ (left~\cite{Chen:2018xof}) and 
$\Delta u(x)-\Delta d(x)$ (right~\cite{Lin:2018qky}) by the LP3 collaboration.
}
\label{fig:lp3}
\end{figure} 
\vspace{-0.15in}

\begin{figure}[!h]
\centering
\begin{minipage}[c]{0.4\textwidth}
\centering
\includegraphics[width=0.82\textwidth]{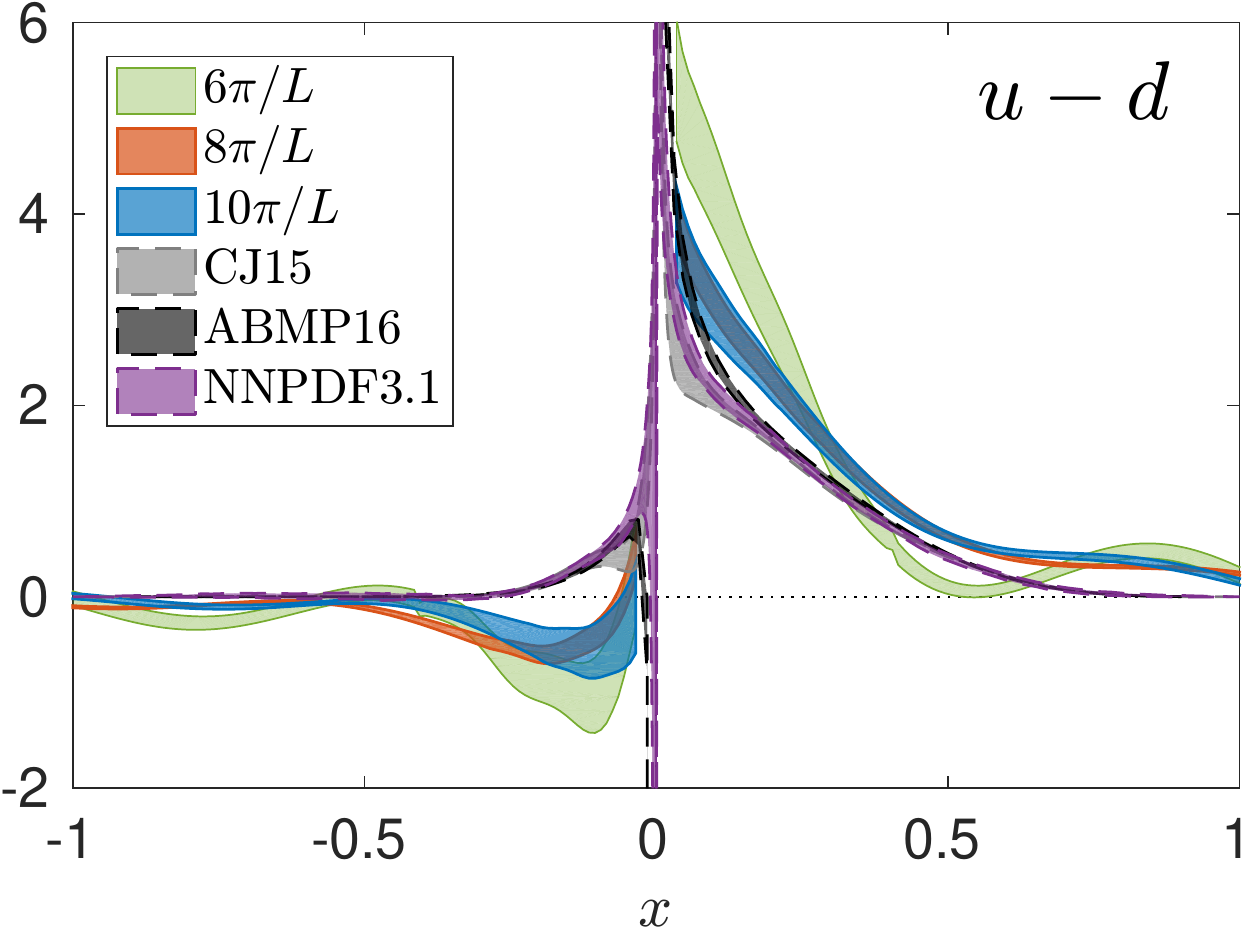}
\end{minipage}
\hskip 0.2in
\begin{minipage}[c]{0.4\textwidth}
\centering
\includegraphics[width=0.82\textwidth]{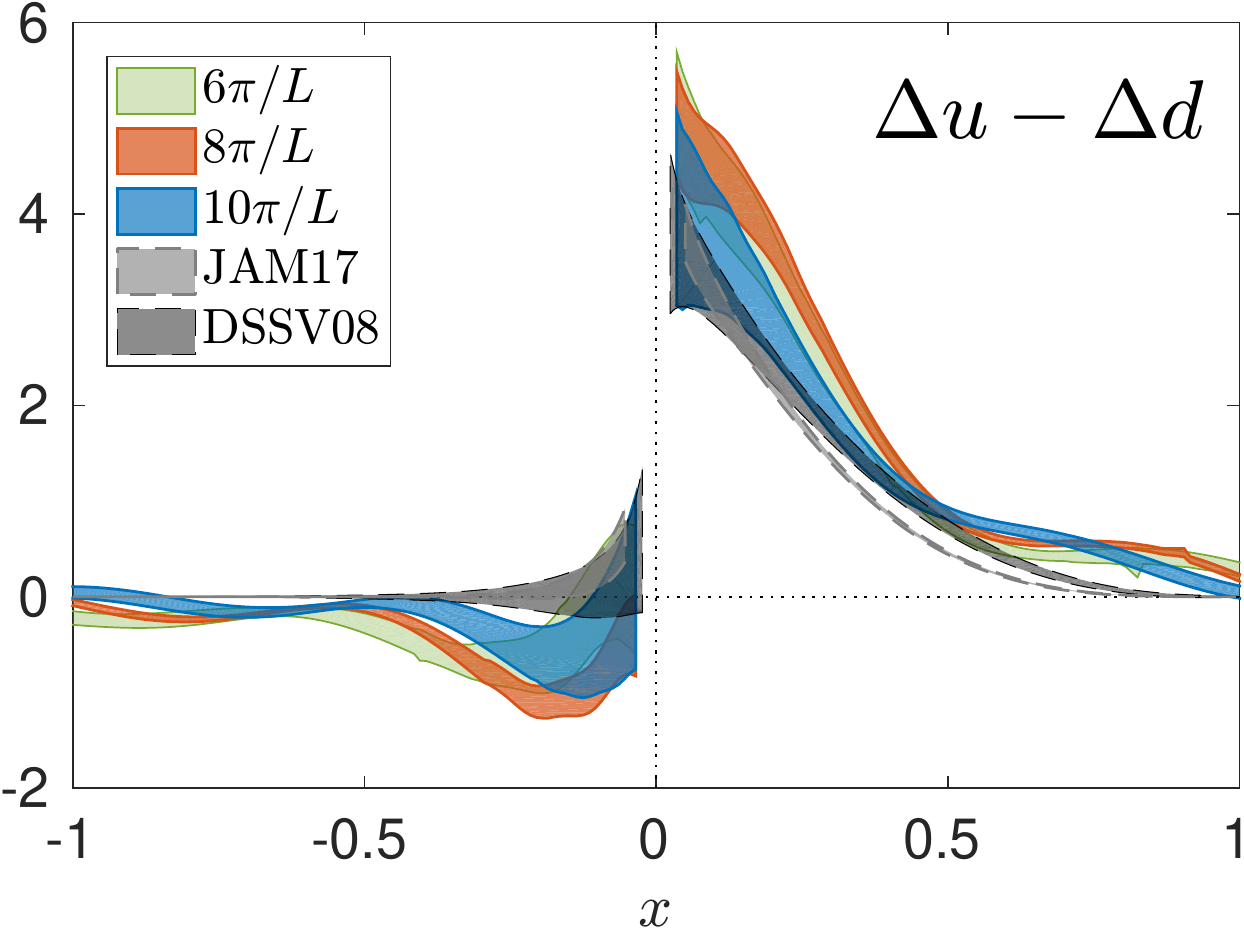}
\end{minipage}
\caption{LQCD determination of $u(x)-d(x)$ (left) and 
$\Delta u(x)-\Delta d(x)$ (right) by the LP3 collaboration~\cite{Alexandrou:2018pbm}.
}
\label{fig:etmc}
\end{figure}

Significant achievements in the LQCD calculations of quasi-PDFs have been made in recent years
\cite{Lin:2017ani,Alexandrou:2018pbm,Chen:2018xof,Alexandrou:2018eet,Lin:2018qky}. At a finite hadron momentum $P_z$, as shown in the right plot in Fig.~\ref{fig:pseudo}, the one-loop matching coefficient plays a very significant role~\cite{Chen:2018xof}.  The latest and the state of art extraction of unpolarized distribution $u(x)-d(x)$, as well as polarized helicity distribution, $\Delta u(x)-\Delta d(x)$, from LP$^3$ collaboration are presented in Fig.~\ref{fig:lp3}, while the same combination of distributions from LQCD calculations of ETMC collaboration are shown in Fig.~\ref{fig:etmc}.  References to many additional LQCD calculations, the current status and challenges for extracting reliable PDFs and other parton correlation functions from such LQCD calculations can be found in Refs.~\cite{Lin:2017snn,Cichy:2018mum}. 

\vspace{-0.1in}
\vspace{-0.15in}

\begin{figure}[!h]
\centering
\begin{minipage}[c]{0.4\textwidth}
\centering
\includegraphics[width=0.98\textwidth]{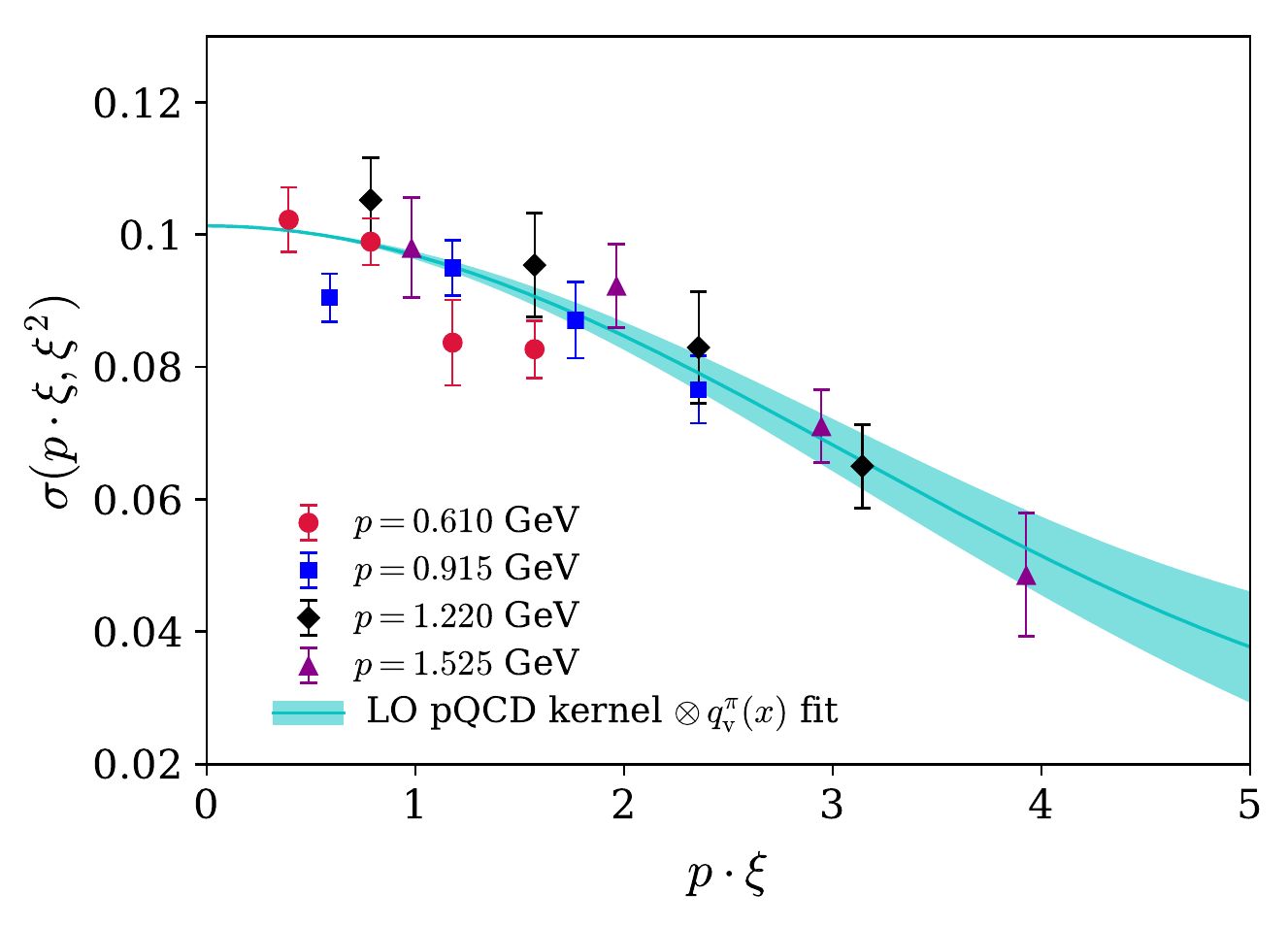}
\end{minipage}
\hskip 0.2in
\begin{minipage}[c]{0.4\textwidth}
\centering
\includegraphics[width=0.98\textwidth]{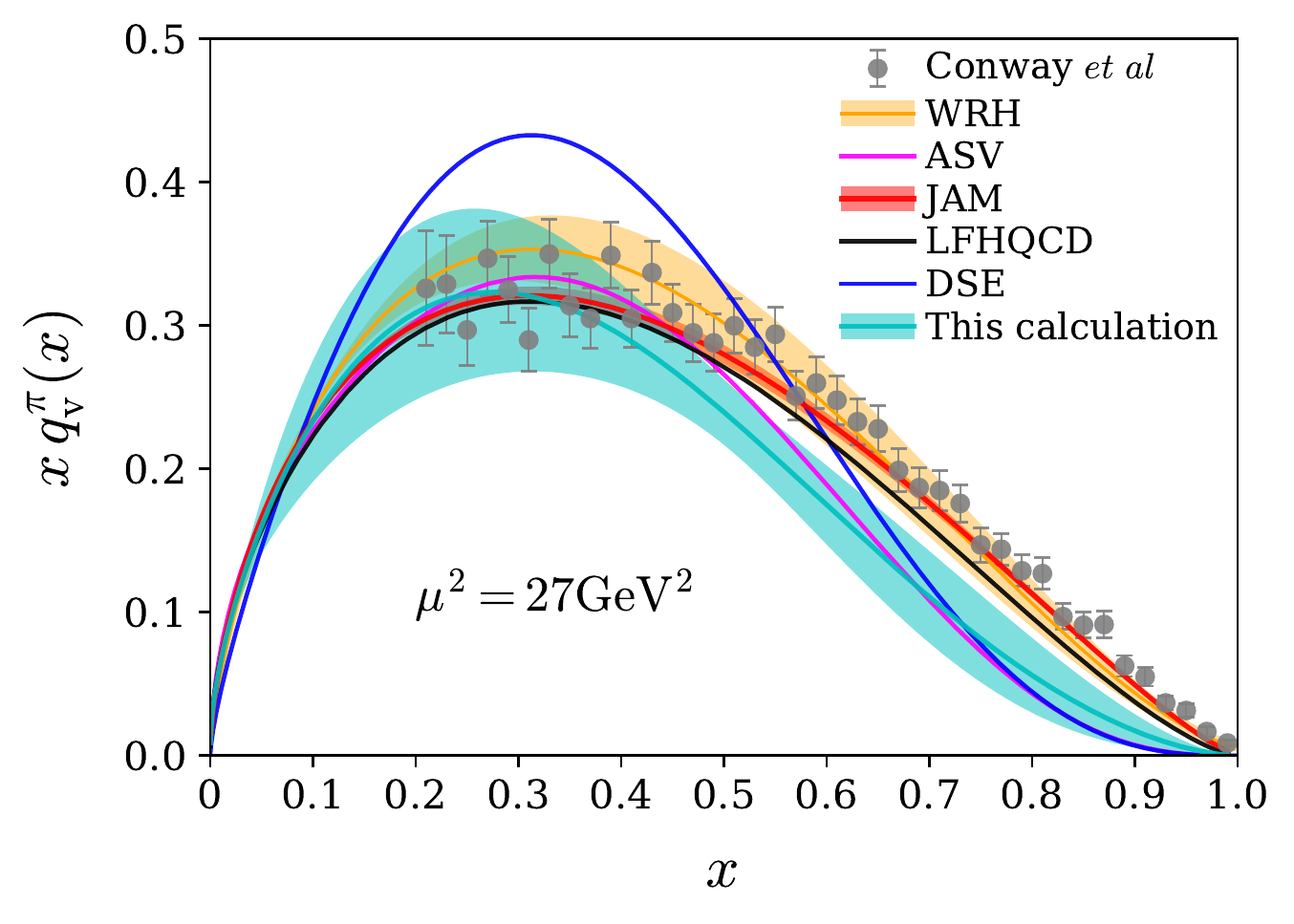}
\end{minipage}
\caption{Left:\ LQCD determination of pion matrix element in Eq.~(\ref{eq:pion})~\cite{Sufian:2019bol}.  Right:\ Comparison of valence quark distribution extracted from LQCD calculation with those obtained with other approaches~\cite{Sufian:2019bol}.
}
\label{fig:pion}
\end{figure}

The good ``lattice cross section'' approach with current-current operators trades the speed of LQCD calculations to a better control of UV renormalization and many more calculable and factorizable observables, by calculating four-point functions in LQCD instead of the three-point functions in the case of calculating quasi- or pseudo-PDFs~\cite{Ma:2017pxb}.  With different currents and QCD symmetries, we could construct various combinations of these currents to have access to different distribution and correlation functions.  For example, under the parity and time-reversal invariance of QCD, pion matrix element of the following combination of a vector current (V) and an axial-vector current (A) has to be antisymmetric in the Lorentz indices of these two currents, 
\begin{equation}
\frac{1}{2}\left[\sigma^{\mu\nu}_{VA}(\xi,p)+\sigma^{\mu\nu}_{AV}(\xi,p)\right]
=\epsilon^{\mu\nu\alpha\beta}\xi_\alpha p_\beta\, T_1(\nu,\xi^2) 
+ \left(p^\mu\xi^\nu-\xi^\mu p^\nu\right) T_2(\nu,\xi^2)\, ,
\label{eq:pion}
\end{equation}
where the LCS, $\sigma_{ij}^{\mu\nu}\equiv\xi^4\langle \pi(p)|J_i^\mu(\xi/2) J_j^\nu(-\xi/2)|\pi(p)\rangle$ with $i,j=V,A$, the Ioffe time, $\omega=p\cdot\xi$, and $T_i$ with $i=1,2$ are functions of Lorentz scalars, $\omega$ and $\xi^2$, which can be factorized to PDFs as~\cite{Ma:2017pxb,Sufian:2019bol},
\begin{equation}
T_i(\omega,\xi^2)=\sum_{a=q,\bar{q},g}
  \int_0^1\frac{dx}{x} f_a(x,\mu^2) C_i^a(x\omega,\xi^2,\mu^2)
  +{\cal O}(\xi^2\Lambda_{\rm QCD}^2)
\label{eq:fac}
\end{equation}
with the lowest order (LO) coefficients $C_1^{(0)}(x\omega,\xi^2)=\frac{2x}{\pi^2}\cos(x\omega)$ and 
$C_2^{(0)}(x\omega,\xi^2)=0$, which gives,
\begin{equation}
\widetilde{T}_1(x,\xi^2)=\int\frac{d\omega}{2\pi} e^{-ix\omega} T_1(\omega,\xi^2) 
\approx \frac{1}{\pi^2}\left[q(x)-\bar{q}(x)\right] = \frac{1}{\pi^2}q_v(x)\, .
\label{eq:c0}
\end{equation}
That is, the LQCD calculated LCS in Eq.~(\ref{eq:pion}) is directly proportional to the pion's valence quark distribution.  In the left plot in Fig.~\ref{fig:pion}, LQCD calculated LCS in Eq.~(\ref{eq:pion}) is shown for several hadron momenta. On the right, the valence quark momentum distribution $xq_v(x)$, extracted from LQCD calculation with the LO kernel is compared with those extracted from data or calculated in models, and the LO result is encouraging.

\vspace{-0.15in}
\vspace{-0.3in}

\section{Summary and outlook}
\label{sec:summary}

I have briefly reviewed what LQCD can and cannot do for calculating the parton distribution and correlation functions, and the new approaches to explore nucleon structure from LQCD calculations.  I presented the recent state-of-art achievements in determine PDFs by the joint effort of LQCD and PQCD communities.  LQCD calculations of hadron structure are complementary to experimental measurements, since LQCD can have access to the kinematic regime, such as the large $x$, where is hard to reach by experimental measurements; and LQCD can study partonic structure of hadrons, such as free neuron, that are hard to do experiments with.  The results of initial exploratory LQCD calculations of parton distribution and correlation functions are very encouraging, while more works are obviously needed.

\vspace{0.2in}
\noindent
{\it Acknowledgement:}\ 
This work is supported in part by the U.S. Department of Energy, Office of Science, Office of Nuclear Physics under Award No.~DE-AC05-06OR23177, within the framework of the TMD Topical Collaboration.



\begin{thebibliography}{9}
\bibitem{Collins:1989gx} 
  J.~C.~Collins, D.~E.~Soper and G.~F.~Sterman,
  {Adv.\ Ser.\ Direct.\ High Energy Phys.\  {\bf 5}, 1 (1989)}
  [{arXiv:hep-ph/0409313}].
  
\bibitem{Brambilla:2014jmp} 
  N.~Brambilla {\it et al.},
  Eur.\ Phys.\ J.\ C {\bf 74}, no. 10, 2981 (2014)
  [{arXiv:1404.3723 [hep-ph]}].

\bibitem{Harland-Lang:2014zoa} 
  L.~A.~Harland-Lang, A.~D.~Martin, P.~Motylinski and R.~S.~Thorne,
  {Eur.\ Phys.\ J.\ C {\bf 75}, 204 (2015)}
  [{arXiv:1412.3989 [hep-ph]}].
  
\bibitem{Dulat:2015mca} 
  S.~Dulat {\it et al.},
  {Phys.\ Rev.\ D {\bf 93},  033006 (2016)}
   [{arXiv:1506.07443 [hep-ph]}].
  
\bibitem{Ball:2017nwa} 
  R.~D.~Ball {\it et al.} (NNPDF Collaboration),
  {Eur.\ Phys.\ J.\ C {\bf 77},  663 (2017)}
  [{arXiv:1706.00428 [hep-ph]}].
  
\bibitem{Alekhin:2017kpj} 
 S.~Alekhin, J.~Bl\"umlein, S.~Moch and R.~Pla\v{c}akyt\.e,
 {Phys.\ Rev.\ D {\bf 96}, 014011 (2017)}
 [arXiv:1701.05838 [hep-ph]].
  
\bibitem{Ethier:2017zbq} 
  J.~J.~Ethier, N.~Sato and W.~Melnitchouk,
  {Phys.\ Rev.\ Lett.\  {\bf 119}, 132001 (2017)}
  [arXiv:1705.05889 [hep-ph]].
  
\bibitem{Lin:2017snn} 
  H.~W.~Lin {\it et al.},
  Prog.\ Part.\ Nucl.\ Phys.\  {\bf 100}, 107 (2018)
  [arXiv:1711.07916 [hep-ph]].
  
\bibitem{Cichy:2018mum} 
  K.~Cichy and M.~Constantinou,
  arXiv:1811.07248 [hep-lat].

\bibitem{Alexandrou:2018sjm} 
  C.~Alexandrou, S.~Bacchio, M.~Constantinou, J.~Finkenrath, K.~Hadjiyiannakou, K.~Jansen, G.~Koutsou and A.~V.~A.~Casco,
  arXiv:1812.10311 [hep-lat], and references therein.
  
\bibitem{Lin:2017stx} 
  H.~W.~Lin, W.~Melnitchouk, A.~Prokudin, N.~Sato and H.~Shows,
  Phys.\ Rev.\ Lett.\  {\bf 120}, 152502 (2018)
  [arXiv:1710.09858 [hep-ph]].
  
\bibitem{Dolgov:2002zm} 
  D.~Dolgov {\it et al.} [LHPC and TXL Collaborations],
  Phys.\ Rev.\ D {\bf 66}, 034506 (2002)
  [hep-lat/0201021].

\bibitem{Gockeler:2004wp} 
  M.~Gockeler {\it et al.} [QCDSF Collaboration],
  Phys.\ Rev.\ D {\bf 71}, 114511 (2005)
  [hep-ph/0410187].
  
\bibitem{Detmold:2001dv} 
  W.~Detmold, W.~Melnitchouk and A.~W.~Thomas,
  Eur.\ Phys.\ J.\ direct {\bf 3}, 13 (2001)
  [hep-lat/0108002].
  
\bibitem{Liu:1993cv} 
  K.~F.~Liu and S.~J.~Dong,
 {Phys.\ Rev.\ Lett.\  {\bf 72}, 1790 (1994)} [{hep-ph/9306299}]
  
\bibitem{Liu:1999ak} 
  K.~F.~Liu,
  {Phys.\ Rev.\ D {\bf 62}, 074501 (2000)} [hep-ph/9910306].
  
\bibitem{Horsley:2012pz} 
  R.~Horsley {\it et al.} (QCDSF-UKQCD Collaboration),
  {Phys.\ Lett.\ B {\bf 714}, 312 (2012)} [{arXiv:1205.6410 [hep-lat]}].

\bibitem{Ji:2013dva} 
  X.~Ji,
  Phys.\ Rev.\ Lett.\  {\bf 110}, 262002 (2013)
  [arXiv:1305.1539 [hep-ph]].
  
\bibitem{Braun:2018brg} 
  V.~M.~Braun, A.~Vladimirov and J.~H.~Zhang,
  Phys.\ Rev.\ D {\bf 99}, 014013 (2019)
  [arXiv:1810.00048 [hep-ph]].
  
\bibitem{Alexandrou:2017huk} 
  C.~Alexandrou, K.~Cichy, M.~Constantinou, K.~Hadjiyiannakou, K.~Jansen, H.~Panagopoulos and F.~Steffens,
  Nucl.\ Phys.\ B {\bf 923}, 394 (2017)
  [arXiv:1706.00265 [hep-lat]].
  
\bibitem{Stewart:2017tvs} 
  I.~W.~Stewart and Y.~Zhao,
  Phys.\ Rev.\ D {\bf 97}, 054512 (2018)
  [arXiv:1709.04933 [hep-ph]].
  
\bibitem{Alexandrou:2019lfo} 
  C.~Alexandrou, K.~Cichy, M.~Constantinou, K.~Hadjiyiannakou, K.~Jansen, A.~Scapellato and F.~Steffens,
  arXiv:1902.00587 [hep-lat].
    
\bibitem{Ji:2017oey} 
  X.~Ji, J.~H.~Zhang and Y.~Zhao,
  Phys.\ Rev.\ Lett.\  {\bf 120}, no. 11, 112001 (2018)
  [arXiv:1706.08962 [hep-ph]].
         
\bibitem{Ishikawa:2017faj} 
  T.~Ishikawa, Y.~Q.~Ma, J.~W.~Qiu and S.~Yoshida,
  Phys.\ Rev.\ D {\bf 96}, no. 9, 094019 (2017)
  [arXiv:1707.03107 [hep-ph]].
  
\bibitem{Green:2017xeu} 
  J.~Green, K.~Jansen and F.~Steffens,
  Phys.\ Rev.\ Lett.\  {\bf 121}, 022004 (2018)
  [arXiv:1707.07152 [hep-lat]].
  
\bibitem{Zhang:2018diq} 
  J.~H.~Zhang, X.~Ji, A.~SchŠfer, W.~Wang and S.~Zhao,
  arXiv:1808.10824 [hep-ph].
  
\bibitem{Li:2018tpe} 
  Z.~Y.~Li, Y.~Q.~Ma and J.~W.~Qiu,
  Phys.\ Rev.\ Lett.\  {\bf 122}, no. 6, 062002 (2019)
  [arXiv:1809.01836 [hep-ph]].
  
\bibitem{Ma:2014jla} 
  Y.~Q.~Ma and J.~W.~Qiu,
  Phys.\ Rev.\ D {\bf 98}, no. 7, 074021 (2018)
  [arXiv:1404.6860 [hep-ph]].
    
\bibitem{Ma:2017pxb} 
  Y.~Q.~Ma and J.~W.~Qiu,
  Phys.\ Rev.\ Lett.\  {\bf 120}, 022003 (2018)
  [arXiv:1709.03018 [hep-ph]].

\bibitem{Radyushkin:2017cyf} 
  A.~V.~Radyushkin,
  {Phys.\ Rev.\ D {\bf 96},  034025 (2017)} [arXiv:1705.01488 [hep-ph]].
  
\bibitem{Orginos:2017kos} 
  K.~Orginos, A.~Radyushkin, J.~Karpie and S.~Zafeiropoulos,
  Phys.\ Rev.\ D {\bf 96}, 094503 (2017)
  [arXiv:1706.05373 [hep-ph]].

\bibitem{Chambers:2017dov}
A.~J. Chambers, R.~Horsley, Y.~Nakamura, H.~Perlt, P.~E.~L. Rakow,
  G.~Schierholz, A.~Schiller, K.~Somfleth, R.~D. Young, and J.~M. Zanotti, 
  {{Phys. Rev. Lett.}{\bf 118} (2017) 242001}
   [arXiv:1703.01153]

\bibitem{Lin:2017ani} 
  H.~W.~Lin {\it et al.} [LP3 Collaboration],
  Phys.\ Rev.\ D {\bf 98}, 054504 (2018)
  [arXiv:1708.05301 [hep-lat]].

\bibitem{Alexandrou:2018pbm} 
  C.~Alexandrou, K.~Cichy, M.~Constantinou, K.~Jansen, A.~Scapellato and F.~Steffens,
  Phys.\ Rev.\ Lett.\  {\bf 121}, 112001 (2018)
  [arXiv:1803.02685 [hep-lat]].
  
\bibitem{Chen:2018xof} 
  J.~W.~Chen, L.~Jin, H.~W.~Lin, Y.~S.~Liu, Y.~B.~Yang, J.~H.~Zhang and Y.~Zhao,
  arXiv:1803.04393 [hep-lat].
  
\bibitem{Alexandrou:2018eet} 
  C.~Alexandrou, K.~Cichy, M.~Constantinou, K.~Jansen, A.~Scapellato and F.~Steffens,
  Phys.\ Rev.\ D {\bf 98}, 091503 (2018)
  [arXiv:1807.00232 [hep-lat]].
  
\bibitem{Lin:2018qky} 
  H.~W.~Lin {\it et al.},
  Phys.\ Rev.\ Lett.\  {\bf 121}, 242003 (2018)
  [arXiv:1807.07431 [hep-lat]].

\bibitem{Sufian:2019bol} 
  R.~S.~Sufian, J.~Karpie, C.~Egerer, K.~Orginos, J.~W.~Qiu and D.~G.~Richards,
  Phys.\ Rev.\ D (in press) arXiv:1901.03921 [hep-lat].
  
\end{thebibliography}
\end{document}